\documentclass[runningheads,a4paper]{llncs}

\usepackage{amssymb}
\usepackage{graphicx}
\usepackage{subfigure}
\usepackage{color}
\usepackage{listings}
\usepackage{textcomp}

\usepackage{url}
\newcommand{\keywords}[1]{\par\addvspace\baselineskip
\noindent\keywordname\enspace\ignorespaces#1}

{\end{list}}

{\end{list}}

{\end{list}}

{\end{list}}

{\end{list}}

{\end{list}}

{\end{list}}

{\end{list}}

{\end{list}}

{\end{list}}

{\end{list}}

{\end{list}}

\def\cocos{CoCoS}

\begin{document}

\mainmatter  

\title{Hallway Monitoring: Distributed Data Processing with Wireless
  Sensor Networks}
\titlerunning{Hallway Monitoring: Distributed data processing with WSNs}

\author{
Tobias Baumgartner
  \and S{\'a}ndor P.\ Fekete
  \and Tom Kamphans
  \and Alexander Kr{\"o}ller
  \and Max Pagel
}
\authorrunning{T. Baumgartner \and S.\,P. Fekete \and T. Kamphans \and
  A. Kr{\"o}ller \and M. Pagel}

\institute{
Braunschweig Institute of Technology, IBR, Algorithms Group,
    Germany\\
  Email:\email{ \{t.baumgartner, s.fekete, t.kamphans, a.kroeller, m.pagel\}@tu-bs.de}
}

\maketitle

\begin{abstract}
  We present a sensor network testbed that monitors a hallway. It
  consists of 120 load sensors and 29 passive infrared sensors (PIRs),
  connected to 30 wireless sensor nodes. There are also 29 LEDs and
  speakers installed, operating as actuators, and enabling a direct
  interaction between the testbed and passers-by. Beyond that, the
  network is heterogeneous, consisting of three different circuit
  boards---each with its specific responsibility. The design of the load
  sensors is of extremely low cost compared to industrial solutions
  and easily transferred to other settings. The network is used for
  in-network data processing algorithms, offering possibilities to
  develop, for instance, distributed target-tracking algorithms. 
  Special features of our installation are highly correlated sensor data
  and the availability of miscellaneous sensor types.
\end{abstract}


\section{Introduction}
\label{sec:intro}

In the research field of wireless sensor networks, a tremendous amount
of fundamental work over the past years has focused on protocol design
and algorithm development. This has led to a high availability of
common routing~\cite{Akkaya2005325},
time-synchronization~\cite{SundararamanBK05},
localization~\cite{Akyildiz02wirelesssensor}, and
clustering~\cite{abbasi07clustering} algorithms---often designed for
general sensor network topologies. Similarly, many
testbeds~\cite{JohnsonSFFSRL06,orbit,motelab} were built during that
time to run these algorithms on real sensor nodes. This became
possible due to both dropping hardware costs and the maturing of
operating systems running on the nodes, simplifying the development
process. Due to the mainly common demands of the algorithms, most of
the available testbeds were also held generic; the main focus was on
the principal functionality of the algorithms and protocols, the aim
being real-world communication behavior and implementations on tiny
micro-controllers.

With the ongoing progress of algorithmic methods and system
technology, it becomes possible as well as important to apply the previously
designed basics to real application areas---thereby often adapting a generic
solution to the
specific needs of a single deployment. Such application areas for
wireless sensor networks are quite dispersed. Deployments vary from
monitoring environmental areas such as volcanos or mountain sides,
over personal area networks in medical applications, to home
automation systems.

Building such real-world applications with actual sensor data
processing is still a challenging task. First, the installation of
specialized sensors often requires a significant amount of additional
work. Second, such sensors may also cost much more than the nodes
themselves---and thus are often not affordable for ordinary sensor
network testbeds.

The design, development, and evaluation of higher-level algorithms in
real deployments in which sensor nodes can share their local knowledge to
obtain global goals requires appropriate sensor data. To carry out
such tests, we developed a hallway monitoring system, consisting of
120 load sensors deployed beneath the hallway floor, and 29 passive
infrared sensors (PIRs) for motion detection. The construction of the
load sensors has already been demonstrated in \cite{bfk-hmsn-09}. The
sensors are connected to nodes, which in turn can then exchange the
measured values. The data is highly correlated, therefore serving as
an ideal testbed for any algorithm performing data aggregation or
in-network data analysis, such as distributed tracking algorithms.

\begin{figure}[htp]
  \centering
    \subfigure[The installation site.]{
      \includegraphics[width=.9\columnwidth]{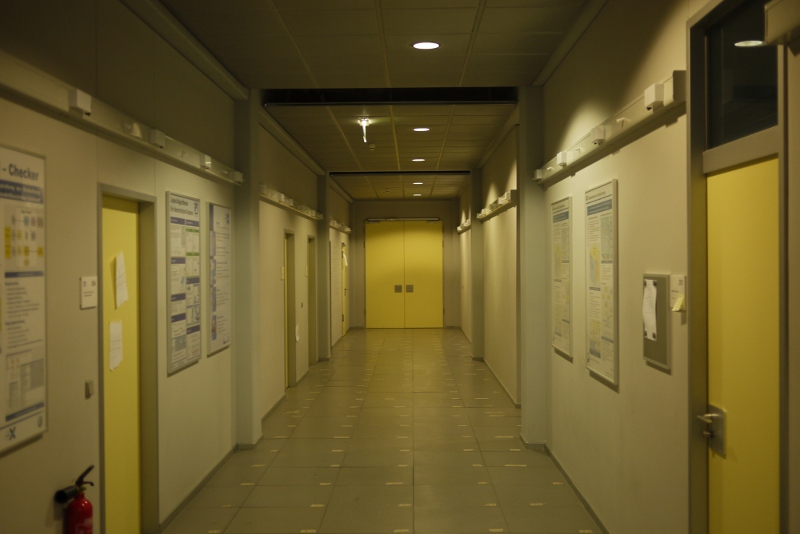}
      \label{fig:whole_hallway}
    }
    \hfill
    \subfigure[Floor tiles rest on columns.]{
      \includegraphics[width=.9\columnwidth]{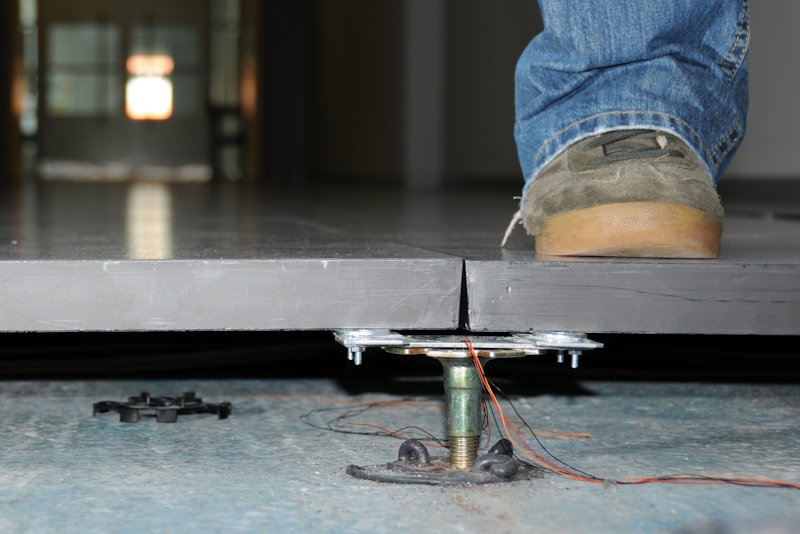}
      \label{fig:floor_plates}
    }
  \caption{Hallway monitoring scenario.}
  \label{fig:hallway_scenario}
\end{figure}

The floor consists of square floor tiles with a side length of 60~cm
each, which are installed on small metal columns. The setup is shown
in \figurename~\ref{fig:hallway_scenario}.

We installed one load sensor on each of these columns. Therefore the
corners of four floor tiles rest on each sensor, and vice versa each
floor tile is monitored by four sensors. Every four load sensors are
connected to a sensor node, which is also installed beneath the
floor. Altogether, the setup consists of 120 load sensors, 29 PIR
sensors, and 30 sensor nodes. The hallway has a width of 3 meters
(corresponding to 5 tiles), and a length of 21.6 meters.

We designed the load sensors ourselves, with a surprisingly cheap
construction. One load sensor costs about 25 Euros---as opposed to
around 200 Euros for industrial manufactured load cells. The lower
price comes with a loss of accuracy, but this loss can be compensated
by sophisticated algorithms for sensor networks, where the nodes do
in-network processing of the highly correlated data.

The rest of the paper is structured as follows. Section
\ref{sec:related} describes similar constructions and related work. In
Section \ref{sec:hallway}, the hallway monitoring system is presented
in detail. In Section \ref{sec:software}, we report on how the sensor
network can be accessed by the public. Section \ref{sec:experiments}
describes first experimental results with the load sensors. We
conclude the paper in Section \ref{sec:conclusion}.


\section{Related Work}
\label{sec:related}

The development of a sensing floor has been proposed by other authors, but
not in the context of a senor network, which is crucial for 
high-level methods and applications.
Addlesee et al. \cite{orl_active_floor} present a design with 3x3
tiles placed on load cells. Similarly, Orr and Abowd
\cite{smart_floor} designed the Smart Floor, also based on load cells.
Neither of the authors considered a sensor network scenario.

While the above approaches make use of expensive load cells, Yiu
and Singh \cite{indoor_tracking} and Kaddoura et
al. \cite{kaddoura_cost-precision_2005} presented designs based on
force sensors. Like in the other cases, there is no distributed data processing,
and the system allowed only for presence detection, as opposed to
more complex information such as the fine-grained resolution of a single step.

Mori et
al.~\cite{mori00-one_room_sensing_system,mori04-people_tracking_floor_pressure}
present both a sensing room with pressure sensors on the floor and
also the furniture, and a sensing floor, which they use to identify
people via their gaits. The latter, gait recognition for people
identification, has also been done by Qian et
al. \cite{qian08-people_identification}. Again, there is no
distributed in-network analysis by small devices like sensor nodes. In
contrast to the previous descriptions, we present a both simple and
highly affordable solution for hallway monitoring. In addition, our
construction allows for the design of sophisticated algorithms running
on tiny sensor nodes.

In general, the possibility of target tracking in indoor environments
is especially interesting in the field of Ambient Assisted Living
(AAL). Elderly, impaired, or disabled people are to be supported by
technical solutions integrated in their homes. Gambi and Spinsante
\cite{aal_multi_camera} present a localization and tracking system
based on multiple cameras. In \cite{performance_accelero}, Jin et
al. analyze the performance of using accelerometers for AAL. However,
having multiple cameras at home comes with a certain discomfort, as
well as the need of constantly wearing sensors when at
home. Our approach can work fully transparent for inhabitants, by
offering the same features as above.


\section{Hallway Construction}
\label{sec:hallway}

We have built a hallway monitoring system in our institute.  To this
end, we designed 120 load sensors, which were installed beneath the
floor tiles in our hallway. A single load sensor and an exemplary
section of the installation are shown in
\figurename~\ref{fig:load_sensor_installation}.  There are also 29 PIR
sensors on the walls to allow combining different kinds of sensor
values in one distributed application. The sensors are connected to a
total of 30 iSense~\cite{buschmann07isense} nodes, which can
communicate over their radio. Finally, there are also actuators
installed---29 light-emitting diodes (LEDs) and speakers to play sound
samples---that are controlled by the sensor nodes, and thus enhance
debugging possibilities of newly designed algorithms. A schematic
diagram of the whole hallway construction with the interconnections of
the several components is shown in \figurename~\ref{fig:hw_structure}.
In the following, each part is described in detail.

\subsection{Load Sensors and PIRs}
\label{ssec:hallway:sensors}

We present a simple---and most notably low-cost---mechanism of
building a single load sensor for our application. We use strain
gauges, which are able to measure minimal strains in the objects to
which they have been glued to. These strain gauges are supplied with a
voltage of a few Volts, whereby they provide an output voltage of just
a few millivolts. Whenever the attached material is strained or
deformed, even by a few nanometers, the output current changes. Such
sensors cost only a few Euros (around 10 Euros apiece in our case).

\begin{figure}[htp]
  \centering
  \subfigure[A single load sensor.]{
    \includegraphics[width=.9\columnwidth]{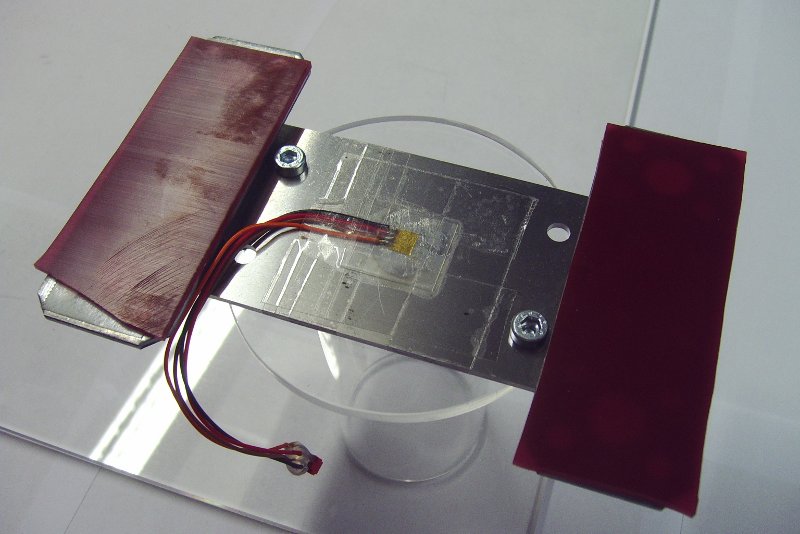}
    \label{fig:load_sensor}
  }
  \hfill
  \subfigure[Load sensors attached to iSense node.]{
    \includegraphics[width=.9\columnwidth]{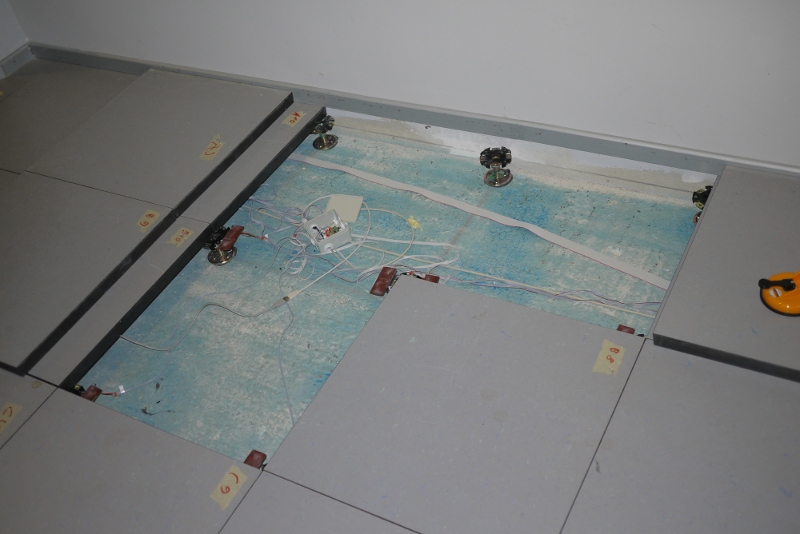}
    \label{fig:floor_tiles}
  }
  \caption{Load sensor installation.}
  \label{fig:load_sensor_installation}
\end{figure}

The strain gauges are attached to small steel plates with a size of
approx. 10x4~cm. We used spring steel for the base construction. The
advantage of spring steel is that it is flexible enough to be strained
by the weight of a person, but also solid enough not to be permanently
deformed. Installing the steel plates under the floor is surprisingly
difficult. Strain gauges measure strains in different directions. If
strain is applied from perpendicular directions, they annihilate each
other, and the sensor does not measure any force. Hence, we enhanced
the construction to deal with this issue. We use two additional steel
plates, each with a spacer. The final construction is shown in
\figurename~\ref{fig:load_sensor}. Like the strain gauges, the
steel construction is surprisingly cheap. We paid around 15 Euros
apiece.

Finally, the load sensors were installed in the hallway. The floor
consists of square floor tiles with a side length of 60~cm each, which
rest on small metal columns. We installed one load sensor on each of
these columns---resulting in a total of 120 sensors beneath the
floor. The sensor data is highly correlated, since the corners of four
floor tiles rest on each sensor, and vice versa each floor tile is
monitored by four sensors. The setup is shown in
\figurename~\ref{fig:floor_tiles}, where three floor tiles were removed
to provide a view of the installation beneath the floor.

In addition to the load sensors beneath the floor, we installed 29 PIR
sensors for motion detection on the walls. Each sensor is placed in a
height of 2.5m, directed approx. 45\textdegree\ downwards. There are
always two sensors face to face with each other, enabling the
observation of a section of the hallway.

This facilitates identifying people by weight, but also by their
motion when passing through the hallway. The different kinds of sensor
values can then be combined to allow for the development of algorithms
for heterogeneous sensor types.

\subsection{Actuators}
\label{ssec:hallway:actuators}

In contrast to the load sensors and PIRs, we have also added actuators
to the hallway. There is a total of 29 lights and speakers installed
on the walls. Each actuator consists of a so-called media board, which
is an extra circuit consisting of an Atmega48, nine LEDs (three red,
three green, three blue) attached to a cooling element, a speaker
connected to the PWM output of the microcontroller, and a 4 GB microSD card for storing sound samples played through the speaker. Each
media board is connected via UART to one sensor node beneath the
floor, and can thus be used as an actuator for direct interaction with
people passing the floor.


\subsection{Sensor Nodes}
\label{ssec:hallway:nodes}

Since the strain gauges have an output of just a few millivolts, they
cannot be measured using ordinary ADCs. We use an additional amplifier
circuit, to which up to six strain gauges can be attached. The circuit
can power the sensors, and also read out and amplify the sensor
output. It bears an Atmega48, which provides multiple ADC ports to
read out the sensor values. The circuit has been designed to be used
directly with our iSense sensor node
platform~\cite{buschmann07isense}, and communicates with the Atmega48
on the amplifier circuit via SPI. Even though it is iterated over up
to six ADCs on the Atmel, and the data is additionally transmitted via
SPI, we achieve a data transfer rate of 800 Hz per load sensor. This
allows for highly fine-grained data-processing, and can lead to
analyzing even single steps of passing people.

\begin{figure}[htp]
  \centering
  \includegraphics[width=.9\columnwidth]{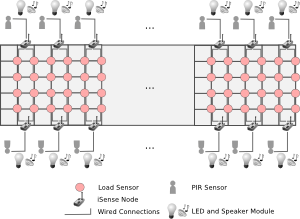}
  \caption{Hallway construction with different kinds of sensors and
    actuators.}
  \label{fig:hw_structure}
\end{figure}

In addition to the connection to the load sensors, the iSense nodes
are also wired to the PIR sensors and actuators on the wall. The whole
setup is shown in \figurename~\ref{fig:hw_structure}. Each wireless
sensor node is connected to four load sensors, one PIR, and one
actuator unit---due to a diverging corridor one wall installation is
missing, resulting in one iSense node without a PIR and LED/speaker
unit connected.

The iSense nodes can then be used for the implementation of high-level
data processing algorithms. For example, by exchanging actual data
over the radio, the nodes can track people walking through the
hallway.

For debugging purposes, the iSense nodes are connected via USB to a
backbone of several PCs. The nodes are powered via this connection. In
addition, they can be re-programmed, and debugging data can be collected
continuously and reliably.


\section{Software Access}
\label{sec:software}

There are two possibilities of accessing the sensor nodes in the
hallway: First, there is an open API offered via web services. Second,
we implemented a Java-based GUI for simple and fast algorithm
development offering a central view on the network.

\subsection{WISEBED API}

The testbed was built in the context of the EU-project
WISEBED~\cite{wisebed}, which aims at the interconnection of different
sensor network testbeds spread over Europe. One goal of the project
is to allow the connection of several testbeds and make them appear as only
one testbed for a user. Moreover, we aim at allowing users to 
connect their own testbeds to one that is part of WISEBED. Therefore, all
APIs that are needed to access a testbed and its sensor nodes are open
to the public. Sensor nodes can be re-programmed, messages can be sent
to the nodes, and debugging output can be collected. All APIs are based on
web services for platform independence.

Since our testbed is part of the WISEBED project, our hallway
monitoring system will be made available for the public---there is, of course, also a user management and
reservation system offered.

\subsection{\cocos~- Java-based GUI}

To facilitate the access to the sensor floor and enable users to
develop their own software, we developed a simple to use Java API
which gives access to the hallway. The so-called ``Corridor Control
System'' (\cocos) consists of a client-server solution which allows
multiple clients to access the floor simultaneously. The server is
embedded as a pluggable module in the WISEBED API and is able to
fully control the floor.

\begin{figure}[htb]
  \centering
  \includegraphics[width=\columnwidth]{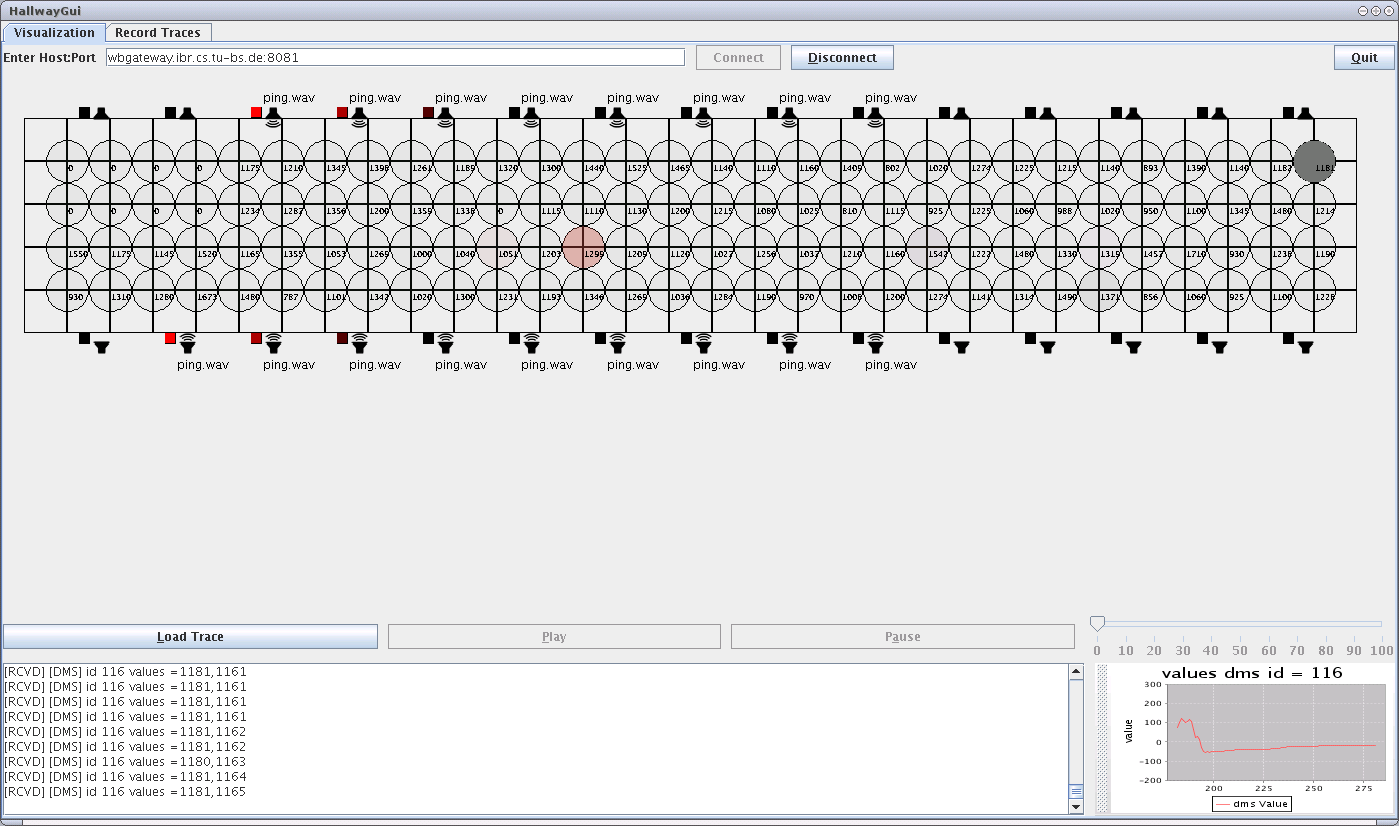}
  \caption{\cocos, a Java-based GUI for accessing the hallway data.}
  \label{fig:cocos}
\end{figure}

\cocos~provides a real-time global view of the sensor floor, which
can be easily accessed to program custom extensions, evaluate sensor
data, or send commands to the sensor floor. Another feature is to
write out sensor data traces, which can be played back later to run
different algorithms on the same data, or work off-line without a
connection to the hallway. The server does not provide a graphical
user interface, but it is possible to connect a GUI-client to the
server via TCP/IP that offers a graphical visualization of the current
floor status, see \figurename~\ref{fig:cocos}. It is also possible to
start an extension from the client to remotely control the corridor,
which makes it possible to work with the testbed from anywhere.

The advantage of \cocos~is that it offers a global view of the whole
network and all sensor data. This simplifies the development process
tremendously, since new ideas can be implemented and evaluated easily
by a centralized algorithm written in Java, and then later translated
to a distributed one working directly on the hallway nodes.


\section{Experimental Study}
\label{sec:experiments}



\begin{figure}[htp]
  \centering
    \subfigure[Data of all sensors.]{
      \includegraphics[width=.9\columnwidth]{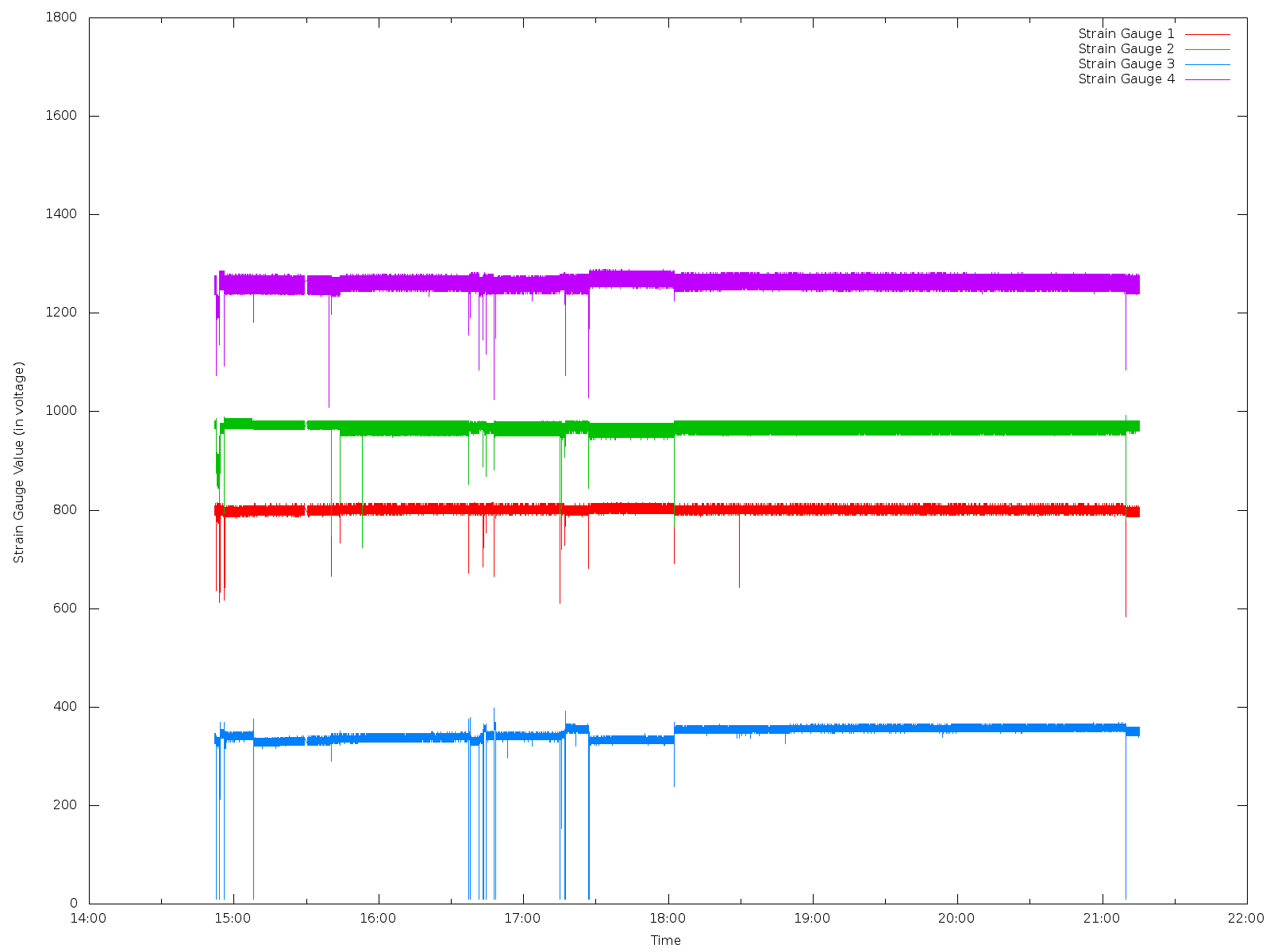}
      \label{fig:four_sensors_hours}
    }
    \hfill
    \subfigure[Detailed view on sensor 1.]{
      \includegraphics[width=.9\columnwidth]{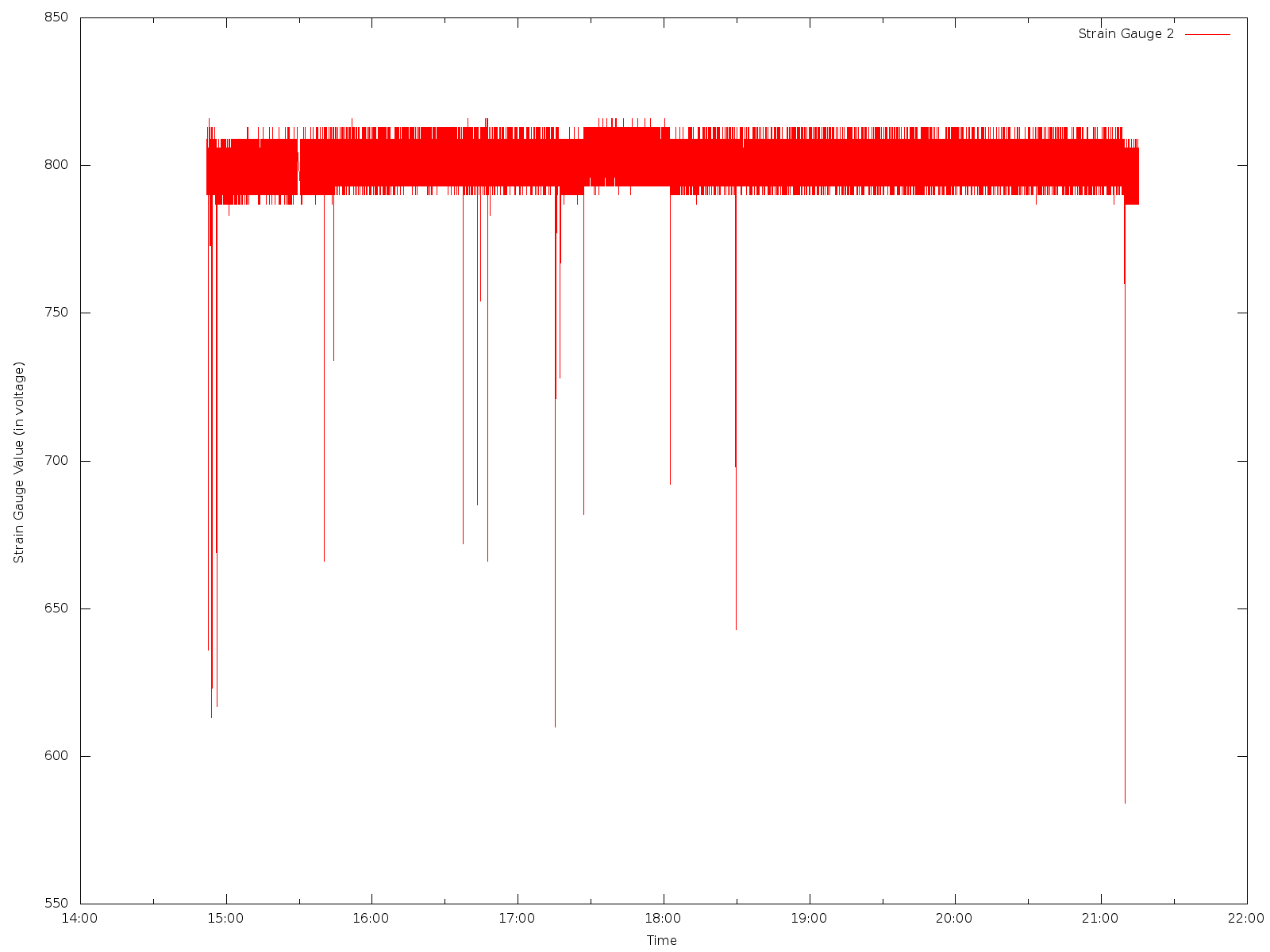}
      \label{fig:sensor1_hours}
    }
    \caption{Data samples of four load sensors installed beneath one
      floor tile.}
  \label{fig:samples_4sens_hours}
\end{figure}

We recorded data of four sensors that are installed beneath one floor
tile to show the correlation of the values, and to have a look into
the data of one sensor in detail. The data was collected with 8~Hz (we
can also take samples with 800~Hz), since it was recorded over several
hours. One data sample is the value read at the ADC of the Atmel
connected to the iSense node, and hence the already amplified strain
gauge value in voltage. When there is no load produced, the sensor
value stays constant. Whenever there is load detected, the value drops
by a certain amount. The results for the four sensors are shown in
\figurename~\ref{fig:four_sensors_hours}.

The zero value of the sensors differs significantly---from around 400
up to approx. 1300. This is due to the self-construction of the
sensors, since the zero value depends on the force the strain gauge is
glued on the steel plate. Analogously, the amplitude is different
from sensor to sensor. Issues arising from differences in zero value or amplitude
can be overcome using appropriate distributed algorithms.
The important observation is the high correlation in the data, which can be seen in
the synchronous amplitude changes of the four sensors.

\figurename~\ref{fig:sensor1_hours} shows the trace of one sensor in
detail. The data is the raw output from the strain gauge, and thus 
basic noise can be seen even when there is no force put on the
sensor. However, one can clearly distinct an amplitude from the noise.

%


\section{Conclusion and Future Work}
\label{sec:conclusion}

We presented a hallway monitoring system based on load and PIR
sensors, which are connected to wireless sensor nodes. The sensor data
is highly correlated, and enables the design of sophisticated
distributed algorithms for target tracking or gait recognition. The
nodes can collaborate to substitute the merely imprecise data of the
load sensors. The inaccuracy of the sensors is outweighed by the
extremely low cost---about 25 Euros per sensor, in contrast to
more than 200 Euros for industrial solutions.

In addition, we also added actuators to the testbed. We installed 29
lights and speakers on the walls to enable the possibility of
interaction between the sensor network and passing people. Both lights
and speakers can be controlled by the sensor nodes beneath the floor,
and can thus be directly integrated in distributed applications.

The whole design also deals with heterogeneity. We have 30 iSense
sensor nodes that can communicate wirelessly. Each node is connected
to a circuit board equipped with an Atmel Atmega48, responsible for
amplifying and receiving the load sensor data. This board is in turn
wired to a circuit board at the walls, controlling the lights and
speakers.

At this point, we finished the construction of the hallway. All sensors are
installed, and the nodes are connected via USB to a backbone for
reliable re-programming and data collection. We have also evaluated
the output of the load sensors.
In the next step, we will develop 
algorithms for more challenging tasks, such as accurate target tracking,
identification of the number of people in the hallway, or the study of
different gaits when the available sample rate of 800 Hz per load
sensor is considered.



\subsubsection*{\ackname}

This work has been partially supported by the European Union under
contract number ICT-2008-224460 (WISEBED). We thank Marcus
Brandenburger, Peter Degenkolbe, Henning Hasemann, Winfried Hellmann,
Bj{\"o}rn Henriks, Roland Hieber, Peter Hoffmann, Hella-Franziska
Hoffmann, Daniel Houschka, Rolf Houschka, Andreas
K{\"o}nig, Christiane Schmidt, Nils Schweer, Stephan Sigg, and
Christian Singer for their assistance in the construction of the
hallway.

\bibliographystyle{abbrv}
\bibliography{main}

\begin{thebibliography}{10}

\bibitem{abbasi07clustering}
A.~A. Abbasi and M.~Younis.
\newblock A survey on clustering algorithms for wireless sensor networks.
\newblock {\em Computer Communications}, 30(14-15):2826--2841, 2007.

\bibitem{orl_active_floor}
M.~Addlesee, A.~Jones, F.~Livesey, and F.~Samaria.
\newblock {The ORL active floor [sensor system]}.
\newblock {\em Personal Communications, IEEE}, 4(5):35--41, Oct 1997.

\bibitem{Akkaya2005325}
K.~Akkaya and M.~Younis.
\newblock A survey on routing protocols for wireless sensor networks.
\newblock {\em Ad Hoc Networks}, 3(3):325 -- 349, 2005.

\bibitem{Akyildiz02wirelesssensor}
I.~F. Akyildiz, W.~Su, Y.~Sankarasubramaniam, and E.~Cayirci.
\newblock Wireless sensor networks: a survey.
\newblock {\em Computer Networks}, 38:393--422, 2002.

\bibitem{bfk-hmsn-09}
T.~Baumgartner, S.~P. Fekete, and A.~Kr{\"o}ller.
\newblock Hallway monitoring with sensor networks.
\newblock In {\em SenSys '09: Proceedings of the 7th ACM conference on Embedded
  network sensor systems}, pages 331--332. ACM, 2009.

\bibitem{buschmann07isense}
C.~Buschmann and D.~Pfisterer.
\newblock {iSense}: A modular hardware and software platform for wireless
  sensor networks.
\newblock Technical report, 6. Fachgespr{\"a}ch Drahtlose Sensornetze der
  GI/ITG-Fachgruppe Kommunikation und Verteilte Systeme, 2007.

\bibitem{aal_multi_camera}
E.~Gambi and S.~Spinsante.
\newblock Multi-camera localization and tracking for ambient assisted living
  applications.
\newblock {\em AALIANCE conference, Malaga, Spain}, Mar 2010.

\bibitem{performance_accelero}
A.~Jin, B.~Yin, G.~Morren, H.~Duric, and R.~Aarts.
\newblock {Performance evaluation of a tri-axial accelerometry-based
  respiration monitoring for ambient assisted living}.
\newblock {\em Engineering in Medicine and Biology Society, 2009. EMBC 2009.
  Annual International Conference of the IEEE}, pages 5677--5680, Sep 2009.

\bibitem{JohnsonSFFSRL06}
D.~Johnson, T.~Stack, R.~Fish, D.~M. Flickinger, L.~Stoller, R.~Ricci, and
  J.~Lepreau.
\newblock Mobile emulab: A robotic wireless and sensor network testbed.
\newblock In {\em INFOCOM}. IEEE, 2006.

\bibitem{kaddoura_cost-precision_2005}
Y.~Kaddoura, J.~King, and A.~S. Helal.
\newblock Cost-precision tradeoffs in unencumbered floor-based indoor location
  tracking.
\newblock In {\em From smart homes to smart care: {ICOST} 2005, 3rd
  International Conference on Smart Homes and Health Telematics}, 2005.

\bibitem{mori00-one_room_sensing_system}
T.~Mori, T.~Sato, K.~Asaki, Y.~Yoshimoto, and Y.~Kishimoto.
\newblock One-room-type sensing system for recognition and accumulation of
  human behavior.
\newblock In {\em IEEE/RSJ International Conference on Intelligent Robots and
  Systems}, 2000.

\bibitem{mori04-people_tracking_floor_pressure}
T.~Mori, Y.~Suemaou, H.~Noguchi, and T.~Sato.
\newblock Multiple people tracking by integrating distributed floor pressure
  sensors and {WID} system.
\newblock In {\em IEEE International Conierence on Systems, Man and
  Cybernetics}, 2004.

\bibitem{smart_floor}
R.~J. Orr and G.~D. Abowd.
\newblock The smart floor: a mechanism for natural user identification and
  tracking.
\newblock In {\em CHI '00: CHI '00 extended abstracts on Human factors in
  computing systems}, pages 275--276, New York, NY, USA, 2000. ACM.

\bibitem{qian08-people_identification}
G.~Qian, J.~Zhang, and A.~Kidan{\'e}.
\newblock People identification using gait via floor pressure sensing and
  analysis.
\newblock In {\em EuroSSC}, pages 83--98, 2008.

\bibitem{orbit}
D.~Raychaudhuri, M.~Ott, and I.~Secker.
\newblock Orbit radio grid tested for evaluation of next-generation wireless
  network protocols.
\newblock In {\em TRIDENTCOM '05: Proceedings of the First International
  Conference on Testbeds and Research Infrastructures for the DEvelopment of
  NeTworks and COMmunities}, pages 308--309, Washington, DC, USA, 2005. IEEE
  Computer Society.

\bibitem{wisebed}
{Seventh Framework Programme FP7 - Information and Communication Technologies}.
\newblock {Wireless Sensor Networks Testbed Project (WISEBED)}, {ongoing
  project since June} 2008.
\newblock {http://www.wisebed.eu}.

\bibitem{SundararamanBK05}
B.~Sundararaman, U.~Buy, and A.~D. Kshemkalyani.
\newblock Clock synchronization for wireless sensor networks: a survey.
\newblock {\em Ad Hoc Networks}, 3(3):281--323, 2005.

\bibitem{motelab}
G.~Werner-Allen, P.~Swieskowski, and M.~Welsh.
\newblock Motelab: a wireless sensor network testbed.
\newblock In {\em IPSN '05: Proceedings of the 4th international symposium on
  Information processing in sensor networks}, page~68, Piscataway, NJ, USA,
  2005. IEEE Press.

\bibitem{indoor_tracking}
C.~Yiu and S.~Singh.
\newblock Tracking people in indoor environments.
\newblock In {\em ICOST'07: Proceedings of the 5th international conference on
  Smart homes and health telematics}, pages 44--53, Berlin, Heidelberg, 2007.
  Springer-Verlag.

\end{thebibliography}

\end{document}